\documentclass[onecolumn,showpacs,preprintnumbers,amsmath,amssymb]{revtex4}
\usepackage{graphics,psfrag}
\usepackage{rotating}
\usepackage[usenames]{color}
\usepackage{amsmath}
\usepackage{bbm}
\usepackage{ulem} 
\newcommand{\be}{\begin{eqnarray}}
\newcommand{\ee}{\end{eqnarray}}

\begin{document}
\large
\title{Chain configurations in light nuclei}

\author{S. Yu. \surname{Torilov}}
\email{torilov@nuclpc1.phys.spbu.ru}
\author{K. A. \surname{Gridnev}}

\affiliation{Department of nuclear physics Saint-Petersburg State
University, Saint-Petersburg, Russia }

\begin{abstract}
The model of nuclear matter built from $\alpha$-particles is
proposed. The strong deformed shape for doubly even N=Z nuclides
from $^{12}$C to $^{24}$Mg has been determined according to this
model.
\end{abstract}
 \maketitle

\section{Introduction}

In spite of considerable efforts to find nuclear chain
configurations, i.e. nuclei consisting of $\alpha$-particle
chains, today it is still not clear if such states exist, except
for the nucleus $^8$Be. Since the end of the sixties this issue
has been widely investigated, both theoretically and
experimentally \cite{a1}. Most of the predictions and experimental
results were related, as a rule, to highly excited states of light
nuclei $^{12}$C, $^{16}$O, $^{20}$Ne and $^{24}$Mg, though there
were attempts to consider heavier nuclei \cite{a2}. Nevertheless,
complications of many-particle problems, which arise due to
inapplicability of the shell model, as well as unverified
interactions in such systems made theoretical predictions
insufficiently accurate. On the other hand, high density of those
light nuclei, which decay predominantly through $\alpha$-channel,
made it hard to compare predictions and the experimental data. In
most cases chain configurations are searched at the levels lying
above the $\alpha$-fragmentation threshold simply because cluster
levels imply energies that exceed the separation energy of one
cluster. In this case the role of the lower chain states is played
by Ikeda diagram \cite{a3}, which presents fragmentation energies
of light nuclei for all possible $\alpha$-cluster states. The most
interesting in this respect is the level 0$_2^+$ in the $^{12}$C
nucleus, which almost exactly matches the energy of decay into
three $\alpha$-particles and decays into $^8$Be and one
$\alpha$-particle. This allows a treating of this level as the
first level in the rotational band formed by the chain
configuration.

\section{Model of binding $\alpha$-particles}

In this paper we undertake another approach, which assumes the
existence of {\it low lying chain configurations}. At the heart of
this approach lies the model of binding $\alpha$-particles
\cite{a31,a32}. The nucleus in this model consists of interacting
$\alpha$-particles, and the sum of interaction energies gives the
binding energy of the nucleus. The coordinates of
$\alpha$-particles in the nucleus are kept frozen, and their
positions are calculated under the following requirements:

1) The binding energy must be at the minimum,

2) The nucleus must have a nearly spherical form.

\noindent The $\alpha$-particle formation proceeds due to the
fluctuations. Inside of the nucleus, the proton and neutron pairs
are correlated dynamically by the proton-neutron interaction Ref.
\cite{a31,a6}. In Fig.~\ref{fig1} is shown the distribution of the
distances between that pairs for the case then this interaction is
very small (i.e. for random distribution of the protons and
neutrons in the nuclear volume). One can see that for light nuclei
we have approximately "equidistant" distribution with RMS~$\sim$~2
fm. In analogy with the Bernal's model of liquid \cite{a5} we
assume that at low excitation energies the structure of
"dynamical" $\alpha$-particles does not change. In this way we
come to the quasi-crystal model of the nucleus, where
$\alpha$-particles form a lattice \cite{a4}. As shown by Hafstad and Teller,
the zero-point energy per bond is quite constant for one
quasi-crystal structure to another \cite{a32}, and this also holds for
four particles in a row \cite{a41}. In Ref. \cite{a7} it was shown that
the interaction energy of $\alpha$-particles can be described by a
wide class of potentials, adequate for atom-atom type of
interactions. We found that the appropriate interaction potential
is in the form of the simplest one-parameter Yukawa form:

\begin{figure}[ht]
\vspace {-8cm}
\centerline{
\includegraphics[width=0.6\textwidth]{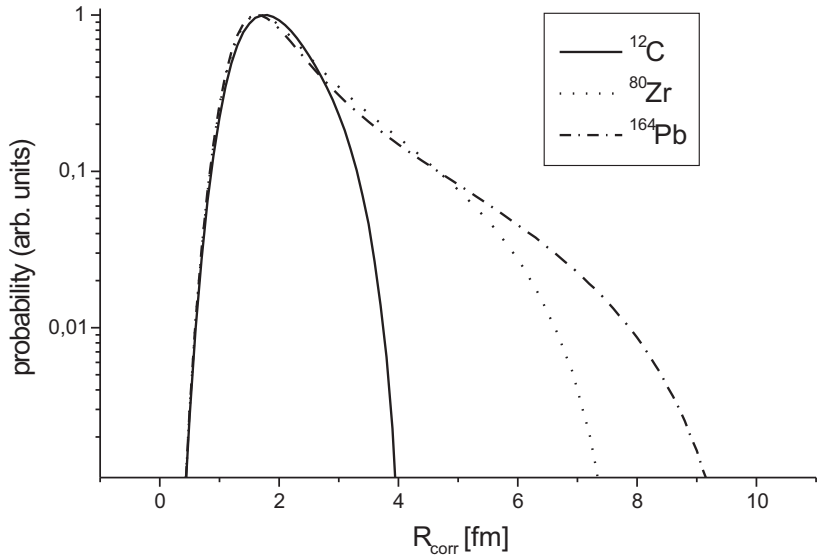}
} \caption{The distribution of the distances between proton and
neutron pairs which form an alpha molecules at a moment. The
approach of the random positions the neutrons and protons where
used.} \label{fig1}
\end{figure}

\renewcommand{\arraystretch}{1.25}
\be
V_{nucl}(r)=-V_0*exp\,(-\gamma r)/r~,
\label{pot}
\ee

\noindent where V$_0$ is fitted through the experimental data
\cite{a7}, $\gamma$ is inverse value of the Compton wavelength of
the neutral $\pi$-meson.

\begin{table}
\caption{Excitation energy in MeV for deformation with $\omega$
N$_\alpha$:N$_\alpha$:1. MBA- model of binding $\alpha$-particles,
LD- liquid drop model Ref.  \cite{a8}, Chain - Energy of the
N$_\alpha$ fragmentation.}
\begin{center}
\begin{tabular*}{0.5\textwidth}{@{\extracolsep{\fill}}lllll}
\hline\hline N$_\alpha$&Nucleus&MBA&LD&Chain
\\
\hline
\rule[-4mm]{0mm}{9mm}3&$^{12}$C&3.0&1.0&7.27
\\
\rule[-4mm]{0mm}{9mm}4&$^{16}$O&8.98&9.0&14.44
\\
\rule[-4mm]{0mm}{9mm}5&$^{20}$Ne&12.99&13.0&19.17
\\
\rule[-4mm]{0mm}{9mm}6&$^{24}$Mg&21.87&20.0&28.48
\\
\hline\hline
\end{tabular*}
\end{center}
\label{tab:prediction}
\end{table}

In Ref.  \cite{a7} was shown that fragmentation energies from this
interaction potential are in a good agreement with experiment for
the light nuclei and qualitatively reflect the behavior of binding
energies up to the stability border.  We are tempting to apply
this model to description of chain configurations, being inspired
by its success in quantum mechanical description of other nuclear
exotics, namely the Bose-Einstein condensation in nuclei
\cite{a7}.

\begin{figure}[ht]
\centerline{
\includegraphics[width=0.6\textwidth]{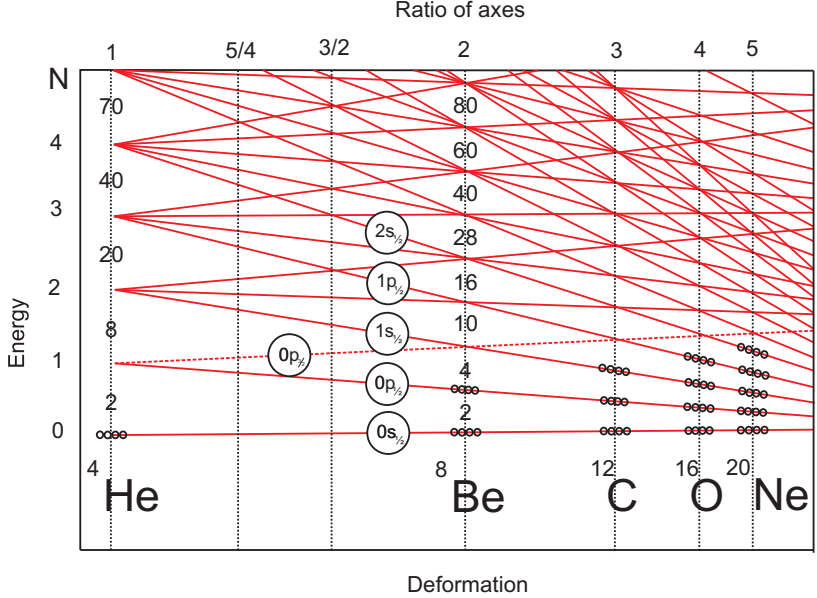}
} \caption{Shells of the deformed oscillator.}
\label{fig2}
\end{figure}

Let us consider the $\alpha$-particle chain, where for interaction
we take the sum of Coulomb and nuclear potentials. Then through
calculating the binding energy E$_b$ in this system we can find
the excitation energy E$^*$, at which the chain may form, from the
change of the number of the bonds and the change in the total
Coulomb energy \cite{a41}. In other words:

\be
E^*=E_{fr}-E_b~,
\label{exc}
\ee

\noindent where E$_{fr}$ is the fragmentation energy. As the
parameters of the model we took the depth of the potential V$_0$ and
the $\alpha$-particle radius R$_\alpha$. We should remark that the
radius value plays a dual role. On one hand it determines the
classical moment of inertia for the $\alpha$-particle. On the other
hand it determines the distance between $\alpha$-particles in the
chain. We have chosen the value R$_\alpha$=1.52~fm, which is close
to the experimental value. The value of V$_0$ (106.7 MeV) was fitted
through fragmentation energies \cite{a7}. Thus, the energy E$^*$ and
the chain's moment of inertia determine the corresponding rotational
band. In Tab. \ref{tab:prediction} the numerical results for E$^*$
from this study are compared with results from work \cite{a8} which
have been determined by the macroscopic-microscopic model. It is
noteworthy that a shape, corresponding to a linear alignment of
$\alpha$-particles, is strongly favored in all double even Z=N
nuclei. In the harmonic-oscillator description this comes about
because the lowest rising levels crosses with the lowest falling
level from the N$^{th}$ shell when the axis ratio is N$_\alpha$:1
(see Fig.~\ref{fig2}). Therefore we can expect some accordance for
these two models. Because the chain rotates around its center of
mass and consists of spinless particles it is invariant with respect
to reflection and hence the band would have positive parity with
even values of the angular momentum. All other levels are
distributed according to the standard rule:

\be
E_J=\frac{J(J+1)\hbar^2}{2\theta}+E^*~,
\label{rot}
\ee

\noindent where J  is angular momentum and $\theta$ is the moment
of inertia. We would like to stress one important point. From the
analysis of the $\alpha$-particles rotational band it follows that
the moment of inertia is not constant and may increase due to the
centrifugal force or decrease due to the reduction in the node
number of the wave function \cite{a9}. From our simple model we
cannot derive the maximum allowed angular momentum or any
restriction on the fluctuations of the moment of inertia. As a
first approximation we assume that the moment of inertia remains
unaltered up to the fragmentation energy. As it was demonstrated
in Ref. \cite{a10}, for the high excitation energy the vibrational
degrees of freedom are important, it gives an analogy with
molecular spectra.

\begin{figure}[ht]
\centerline{
\includegraphics[width=0.6\textwidth]{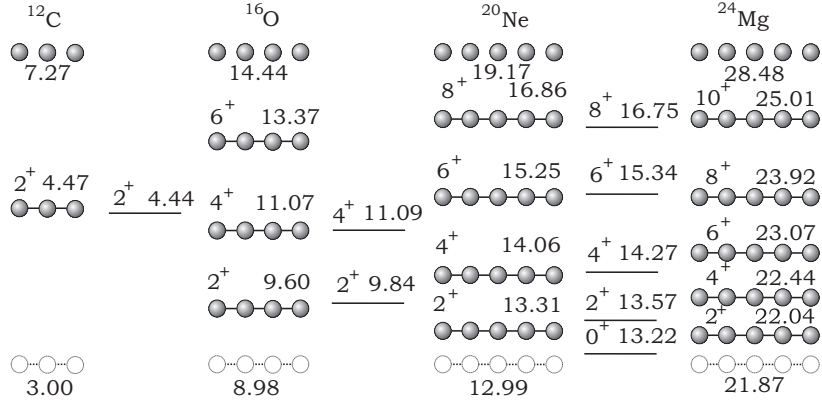}
} \caption{Theoretical (spheres) and experimental (solid line)
energies of rotational bands of chain configurations in nuclei
$^{12}$C, $^{16}$O, $^{20}$Ne and $^{24}$Mg (the theory only).
Below are the values of zeros for the first levels of rotational
bands, upperline corresponds to the fragmentation energy.}
\label{fig3}
\end{figure}

Now, to calculate the levels of the rotational band we should know
the moment of inertia of the chain configuration. This is
calculated according to classical mechanics:

\be
\theta_N=\sum_{i=1}^NM_{\alpha}a_i^2+\frac{2}{5}NM_{\alpha}R_{\alpha}^2~,
\label{rot1} \ee

\noindent where N is the number of $\alpha$-particles, M$_\alpha$
is the $\alpha$-particle mass and a$_i$ are the distances from
$\alpha$-particles to the center of mass of the chain
configuration.

Fig.~\ref{fig3} shows the obtained energy values for rotational
bands in the interval E$^*\le$E$_J\le$E$_{fr}$. One finds a good
agreement with the experimental values, which have the spin values
"required" by the model. The first point to note is that there are
no experimental candidates for the predicted 0$^+$ levels in
$^{12}$C and $^{16}$O. As it is well known that the 0$^+_2$ state in
$^{12}$C (Hoyle state) has an excitation energy of 7.65 MeV and
there is no evidence of low laying 0$^+$ level at 3 MeV. Anywhere
calculation predict for this nucleus an oblate ground state and
prolate excited state \cite{a8} in agreement with our results. In
$^{16}$O the predicted level possible has the same energy as 2$^-$
(8.87 MeV) state. A substantial deviation happens for the level
6$^+$ in $^{16}$O. This could be explained as follows. As noted
above, in this model we cannot give an exact quantum mechanical
value of the maximum momentum as it is usually done with the help of
the Wildermuth condition. Yet, one can determine the maximum
angular momentum from classical considerations. The angular
frequency of the rotating deformed nucleus is

\be
\omega=\frac{\hbar(J(J+1))^{1/2}}{\theta}~.
\label{omega}
\ee

\begin{figure}[t!]
\centerline{
\includegraphics[width=0.4\textwidth]{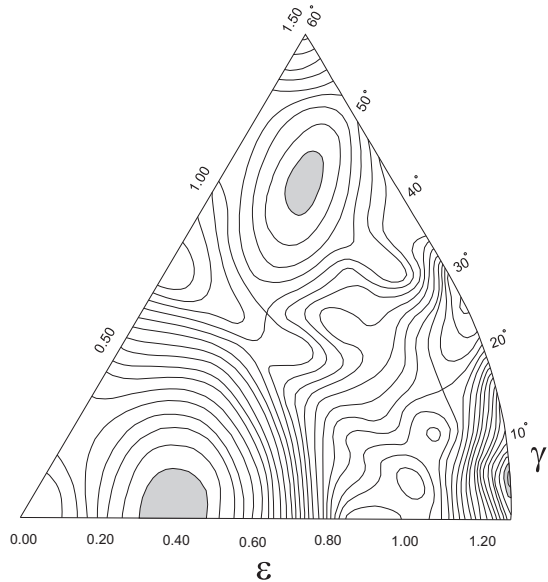}
} \caption{The potential-energy surface for $^{20}$Ne Ref.  \cite{a8}.}
\label{fig4}
\end{figure}

\begin{figure}[ht]
\centerline{
\begin {sideways}
\includegraphics[width=0.4\textwidth]{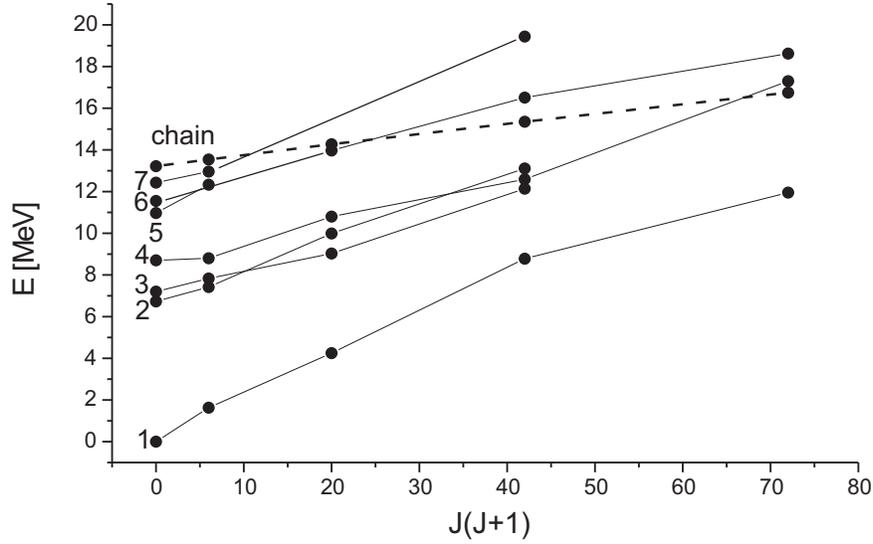}
\end {sideways}
} \caption{The rotational bands for $^{20}$Ne Ref. \cite{a14}.}
\label{fig5}
\end{figure}

When one calculates the values of $\omega$ for considered nuclei,
one finds that the frequencies corresponding to maximal angular
moments are nearly equal to 0.8-1.0 MeV. Thus the adiabaticity
condition in this model  $\omega_{rot}\ll \omega_{osc}$ is
fulfilled to a reasonably good extent and a higher angular
momentum would break it. As was shown in \cite{a12} the maximal
angular momentum of the hyperdeformed nucleus $^{152}$Dy is around
100 $\hbar$. According to \cite{a13} this value is in good
agreement with drop model's prediction for rotation nuclei. The
radius of nuclei is proportional to A$^{1/3}$, so the same
velocity on the surface leads to maximal angular momentum for
nuclei $^{12}$C, $^{16}$O and $^{20}$Ne -3,5 and 7 which is in
agreement with the experiment. Unfortunately, it is impossible to
calculate the forces within this model directly because we use the
model of a hard rotator, which is obviously invalid for higher
angular momenta. Nevertheless, the forces resulting from the above
potential are close by the order of magnitude to proclaimed
values. In principle this model is capable of reproducing not only
rotational levels of the chain configurations but all cluster
levels in the Ikeda diagram, for example, for the rotational band
of $^{16}$O. The excitation energy of two protons and two neutrons
from the {\it p}-shell to the {\it sd}-shell gives rise to a very
large triaxial deformation. As was shown in \cite{a15} this oblate
shape is similar to the C$_4$-symmetric shape (four
$\alpha$-particles are placed in a plane forming a square). For
this shape the calculation gives the excitation energy of around 6
MeV and the parameter $\hbar^2/2\theta$ of around 0.22 MeV, so we
have a good agreement with experimental data. A too low excitation
energy of chain configurations alerts, when one compares it with
fragmentation energy. However, one should keep in mind that
sometimes the $\alpha$-particle bands in light and intermediate
nuclei start deeply below the threshold of one $\alpha$-particle
separation, i.e. below the corresponding value on the Ikeda
diagram. The most interesting from this point of view is the
nucleus of $^{20}$Ne. In Tab. \ref{tab:numbers} there are the
parameters of the levels \cite{a14}, which are possible form the
rotational band. Like for $^{24}$Mg rotational band in \cite{a11}
we can consider the potential-energy surface for $^{20}$Ne. We can
see \cite{a8} that really there are some minima in this potential
surface and one of these have strong prolate deformation (see
Fig.~\ref{fig4}). For this nucleus every 0$^+$ state correspond to
the rotational band (see table 20.20 in \cite{a14}). Therefore, as
is shown in Fig.~\ref{fig5} we have 7 rotational bands and some of
them are described in terms of liquid drop model and correspond to
the locals minima on Fig.~\ref{fig4}. The rotational band which
was discussed above corresponds to 0$^+_8$ and has J(J+1)
dependence. In conclusion, we would like to note that these
predictions are need in the experimental confirmation.

\begin{table}[h!]
\caption{Rotational band K=0$^+_8$ in $^{20}$Ne.}
\begin{center}
\begin{tabular*}{0.4\textwidth}{@{\extracolsep{\fill}}lll}
\hline\hline J&E$^*$ [MeV]&$\Gamma$ [keV]
\\
\hline
0&13.22&40
\\
2&13.53/13.57&61/12
\\
4&14.27&92
\\
6&15.35&-
\\
8&16.75&160
\\
\hline\hline

\end{tabular*}
\end{center}
\label{tab:numbers}
\end{table}

\section*{Acknowledgments}

The work was done under the support of the Russian Federation
President's Grant ü MK-1897.2005.2

\end{document}